\documentclass[twocolumn,showpacs,preprintnumbers,amsmath,amssymb,superscriptaddress,prb]{revtex4}

\usepackage{graphicx}
\usepackage{dcolumn}
\usepackage{bm}
\usepackage{textcomp}
\usepackage{hyperref}
\usepackage{color}

\newcommand{\rung}{\mbox{\scriptsize rung}}

\begin{document}

\title{Dynamics of doublon-holon pairs in Hubbard two-leg ladders}

\author{Luis G.~G.~V. Dias da Silva}%
\affiliation{Instituto de Física, Universidade de S\~ao Paulo, C.P.
66318, S\~ao Paulo, SP, Brazil, 05315-970.}

\author{G. Alvarez}
\affiliation{Computer Science \& Mathematics %
Division and Center for Nanophase Materials Sciences, Oak Ridge National Laboratory,%
 \mbox{Oak Ridge, TN 37831}, USA}

\author{E. Dagotto}
\affiliation{%
Department of Physics and Astronomy, University of Tennessee, Knoxville, Tennessee 37996, USA, and %
Materials Science and Technology Division, Oak Ridge National Laboratory, Oak Ridge, Tennessee 37831, USA}

\date{\today}

\begin{abstract}

The dynamics of holon-doublon pairs is studied in Hubbard two-leg ladders using
the time-dependent Density Matrix Renormalization Group method. We find that
the geometry of the two-leg ladder, that is qualitatively different from a
one-dimensional chain due to the presence of a spin-gap, strongly affects the
propagation of a doublon-holon pair. Two distinct regimes are identified. For
weak inter-leg coupling, the results are qualitatively similar to the case of
the propagation previously reported in Hubbard chains, with only a
renormalization of parameters. More interesting is the case of strong inter-leg
coupling where substantial differences arise, particularly regarding the double
occupancy and properties of the excitations such as the doublon speed. Our
results suggest a connection between the presence of a spin gap and qualitative
changes in the doublon speed, indicating a weak coupling between the doublon
and the magnetic excitations.

\end{abstract}





\maketitle

\section{\label{sec:intro}Introduction}

Quasi-one dimensional correlated quantum systems have received considerable
attention from the Condensed Matter community because they display exotic
properties that arise from the dimensional confinement of electrons. A famous
example are the quantum spin systems with the geometry of ``ladders'', namely
infinitely long one-dimensional ``legs'' that are coupled along ``rungs'' via
tight binding hopping terms with strengths comparable or larger than those
along the legs.\cite{Dagotto:Rep.Prog.Phys.:1525:1999} Previous research has
shown that ladders with an odd number of legs behave similarly as truly
one-dimensional systems in the sense that spin-spin correlations decay with
distance following power laws. However, for even number of legs (two in
particular) the decay is exponential due to the presence of a so-called spin
gap.\cite{Dagotto:Phys.Rev.B:5744:1992,White:Phys.Rev.Lett.:886:1994,Martins:Phys.Rev.Lett.:3563:1997,Laukamp:Phys.Rev.B:10755:1998}
The existence of this spin gap has been confirmed experimentally in materials
such as the copper oxide SrCu$_2$O$_3$. For the case of two-leg ladders, the
spin gap is caused by the dominance of spin singlet configurations along the
rungs, which have a spin gap between singlet and triplet states. The existence
of a ground state dominated by spin singlet arrangements resembling resonant
valence bond states, and other properties such as the prediction of
superconductivity upon doping, have established interesting analogies between
two-leg ladders and the two-dimensional high temperature superconductors based
on CuO$_2$ layers.\cite{Rice:EPL:445:1993} Two-leg ladders can also be found in
the context of organic compounds such as BPCB ((C$_5$ H$_{12}$ N)$_2$ Cu
Br$_4$),\cite{Bouillot:54407:2011} showing that the importance of two-leg
ladders is not only restricted by its similarity with the high-$T_c$ cuprates
but it is broader, covering several families of materials.
\begin{figure}
\centering{
\includegraphics[width=0.9\columnwidth]{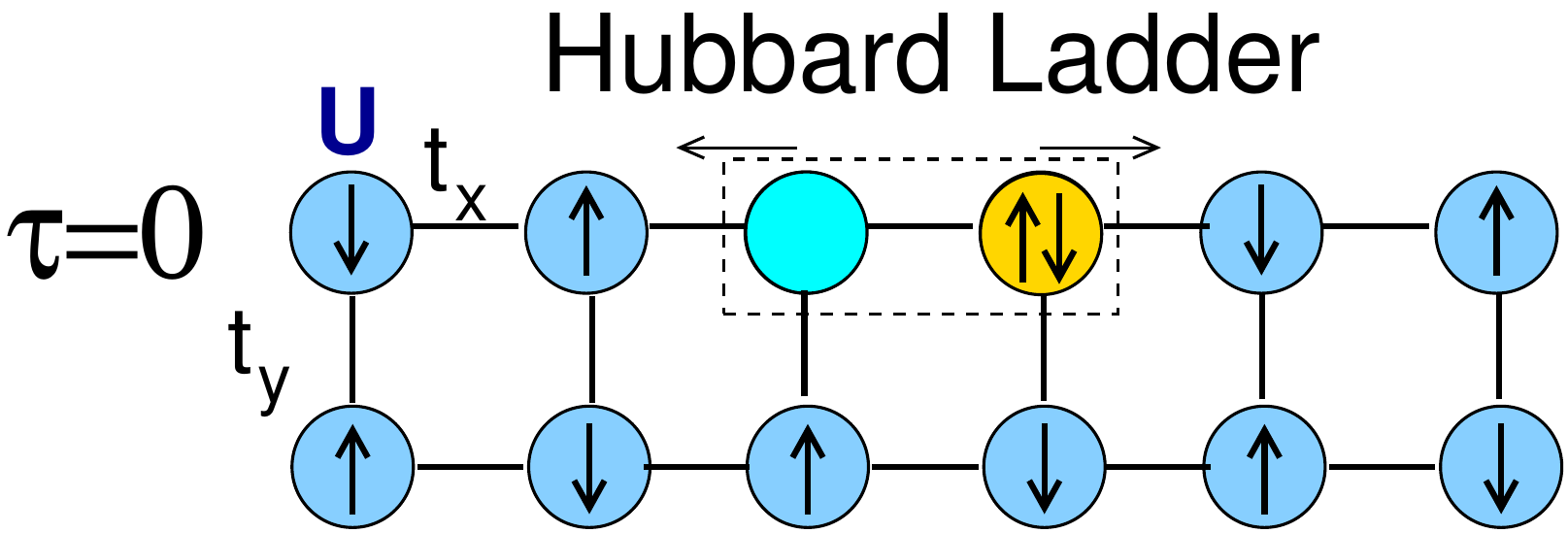}
} \caption{\label{fig:ladder} (Color online) Creation of a doublon-holon pair
in a two-leg Hubbard ladder.}
\end{figure}

The studies of two-leg ladders are not restricted to spin systems, but
considerable work has been devoted to the exploration of both the spin and the
charge degrees of freedom in the context of Hubbard models defined on two-leg
ladder geometries. In fact, computational results for Hubbard ladders with
strong rung (or inter-leg) interaction show the presence of a spin gap for all
interaction
strengths.\cite{Noack:EPL:163:1995,Noack:PhysicaC:281:1996,Noack:Phys.Rev.B:7162:1997}
 However, not much is known about the real time dynamics
of charge excitations in ladder systems. Given the prospect that highly
correlated oxides can become technologically useful for light-to-energy
conversion, the understanding of this interplay is very important. A related
issue is the role of charge excitations in the transport properties of strongly
correlated nanostructures, as highlighted by recent experiments.
\cite{Wei:Nanoscale:3509-3521:2011}

A crucial question is whether charge excitations in a correlated insulator will
be able to properly transfer the charge into the metallic contacts, thus
establishing a steady-state photocurrent. This has been a motivating factor for
studies of exciton-like pairs of charge excitations in Mott insulators
\cite{Al-Hassanieh:166403:2008,DiasdaSilva:125113:2010,Sensarma:35303:2009}
and, more recently, in bilayer
systems.\cite{Rademaker:EPL:27004:2012,Rademaker:NewJournalofPhysics:83040:2012}
In Hubbard systems, the elementary charge excitation that preserves the total
number of electrons is a ``doublon-holon" pair, formed by a double-occupied
site (doublon) and an ``empty" one (holon), as shown schematically in
Fig.~\ref{fig:ladder}. In some situations, the holon and doublon can attract
each other forming a ``Hubbard exciton", as revealed through spectroscopic
evidence by a recent experiment in the transition metal oxide YVO$_3$.
\cite{Novelli:arXiv:1205.4609::2012}

Thus, it is important to understand how interactions, quantum fluctuations, and
dimensionality can affect the lifetime of such charge excitations. In our
previous effort,\cite{Al-Hassanieh:166403:2008} it was shown that the mechanism
for holon-doublon decay into magnetic excitations is inefficient in strictly
one-dimensional Hubbard chains. In the present paper, we extend these ideas to
investigate the dynamics of doublon-holon pairs in Hubbard two-leg ladders,
which are qualitatively different from chains due to the presence of a spin
gap. We find that the geometry of the ladder strongly affects the propagation
of a doublon-holon pair and identify two distinct parameter regimes.

For weak inter-leg coupling, the results are qualitatively similar to the case
of the chain \cite{Al-Hassanieh:166403:2008,DiasdaSilva:125113:2010}, with some
renormalization of the parameters. More interestingly is the case of strong
inter-leg coupling: in this regime, that is drastically different from a
one-dimensional chain, quantities such as the double occupancy and the
properties of the excitations (e.g., the doublon speed) are qualitatively
affected by the ladder geometry.

The paper is organized as follows: the Hubbard ladder model and the specific
technical details of the tDMRG method used are presented in
Sec.~\ref{sec:modelandmethods} and in Appendix ~\ref{sec:TechnicalComments}. In
Sec.~\ref{sec:results}, we present the results for the charge
(Sec.~\ref{sec:chargedynamics}) and double occupancy dynamics
(Sec.~\ref{sec:doubleoccupancy}) as well as the dependency of the excitation
speed with the spin-gap model (Sec.~\ref{sec:doublonspeed}). Our concluding
remarks are presented in Sec.~\ref{sec:conclusion}.

\section{\label{sec:modelandmethods}Model and Methods}

The Hubbard ladder Hamiltonian is written as
\begin{equation}
\hat{H}=\sum_{i,j}t_{ij}\hat{c}^\dagger_{i\sigma}\hat{c}_{i\sigma} +U\sum_i
\hat{n}_{i\uparrow}\hat{n}_{i\downarrow} \; ,
\label{Eq:Hamiltonian}
\end{equation}
where the notation is similar to that of
Ref.~\onlinecite{Al-Hassanieh:166403:2008}. The hopping matrix $t$, however,
now corresponds to that of a two-leg ladder, such that $t_{ij}=t_x$ if $i$ and
$j$ are nearest neighbors along the $x$ direction (i.e., the leg direction),
$t_{ij}=t_y$ if $i$ and $j$ are nearest neighbors along the $y$ direction, and
$0$ (zero) otherwise. In the following, we set $t_x=1$ and use it as our energy
unit.

In order to create charge excitations, we define holon- and doublon-creation
operators as $h^\dagger_i = \sum_\sigma
\hat{c}_{i\sigma}(1-\hat{n}_{i\bar{\sigma}})$, and $d^\dagger_i = \sum_\sigma
\hat{c}^\dagger_{i\sigma}\hat{n}_{i\bar{\sigma}}$ respectively, where
$\bar{\uparrow}=\downarrow$.
As in Ref.~\onlinecite{Al-Hassanieh:166403:2008}, we model the electron- and
hole-like excitations (created by light absorption, for instance) as an excited
state $|\Psi_e\rangle=h^\dagger_i d^\dagger_j|\Psi_0\rangle$, where
$|\Psi_0\rangle$ is the ground state of the Hubbard-ladder system at
half-filling, and $i$ and $j$ are chosen fixed sites where the excitation
occurs. For concreteness, as depicted in Figure \ref{fig:ladder}, we choose
these sites to be the central sites of the upper leg, which allows a direct
comparison to the results depicted in
Refs.~\onlinecite{Al-Hassanieh:166403:2008,DiasdaSilva:125113:2010} for the
case $t_y=0$.

Next comes the time evolution of the excited state with the tDMRG method.
\cite{PhysRevLett.93.076401,DaleyJSM2004} In order to time-evolve the system,
we will use the time-step-targetting Krylov method
(Ref.~\onlinecite{re:manmana05} and references therein) and follow the
implementation found in Ref.~\onlinecite{Alvarez:56706:2011}. We find this
approach to be better suited for generic Hamiltonians in the ladder geometry
rather than the Suzuki-Trotter decomposition method, which would require the
ladder to be treated as a series of coupled dimers.\cite{Bouillot:54407:2011}
We have also compared some results to a Runge-Kutta approach and found
excellent agreement.

We calculate the state $|\Psi (\tau)\rangle=e^{-i \hat{H} \tau}
|\Psi_e\rangle$, with $\hat{H}$ given by Eq. (\ref{Eq:Hamiltonian}), as a
function of time $\tau$ (such that $|\Psi (0^{+})\rangle=|\Psi_e\rangle$) and
then compute real observables as described in Sec.~\ref{sec:results}. Unless
otherwise stated, the results shown are for a 20-site ($10 \times 2$) ladder at
half-filling, with time step $\Delta \tau=0.2$ and times up to $\tau \leq 6$
(in units of $\hbar/t_x$). As we discuss in Appendix
\ref{sec:TechnicalComments}, accumulated errors during the time evolution
introduce a constraint in the accuracy of the results at long time scales. For
this reason, we have opted to show results with times only up to
$\tau_{\mbox{\scriptsize max}}=6$ (see Appendix \ref{sec:TechnicalComments} for
details).

\section{Results}
\label{sec:results}

In this section, we show results for the local charge and double occupancy at
each site, defined respectively as
\begin{eqnarray}
\langle n_{i}\rangle(\tau) &=&\sum_{\sigma=\uparrow, \downarrow}\langle\Psi (\tau)|\hat{c}^{\dagger}_{i \sigma} \hat{c}_{i \sigma} |\Psi (\tau) \rangle/\langle \Psi (\tau) |\Psi (\tau) \rangle \; ,\nonumber \\
\langle n^d_{i}\rangle(\tau)&=&\langle\Psi (\tau)|\hat{n}_{i
\uparrow}\hat{n}_{i \downarrow}|\Psi (\tau) \rangle/\langle \Psi (\tau) |\Psi
(\tau) \rangle \; , \label{Eq:chargedoccdef}
\end{eqnarray}
using the state $|\Psi (\tau)\rangle=e^{-i H \tau} |\Psi (0^{+})\rangle$. The
calculations are performed for a 20-site ($10 \times 2$) ladder at half-filling
unless otherwise noted. In our convention, the sites are labeled as follows:
the leftmost upper site is site 0, the leftmost lower site is site 1, and the
site index increases by one in the vertical top-down direction and by two in
the horizontal-right direction. Thus, even numbered sites (starting at ``0")
are located in the upper leg and odd-numbered sites are in the lower leg. The
doublon-holon pair was created at time $\tau=0^+$ in the center sites of the
upper leg, indicated as sites ``8" (holon) and ``10" (doublon).

\subsection{Charge dynamics}
\label{sec:chargedynamics}

\begin{figure}[t]
\includegraphics[width=1.0\columnwidth,clip]{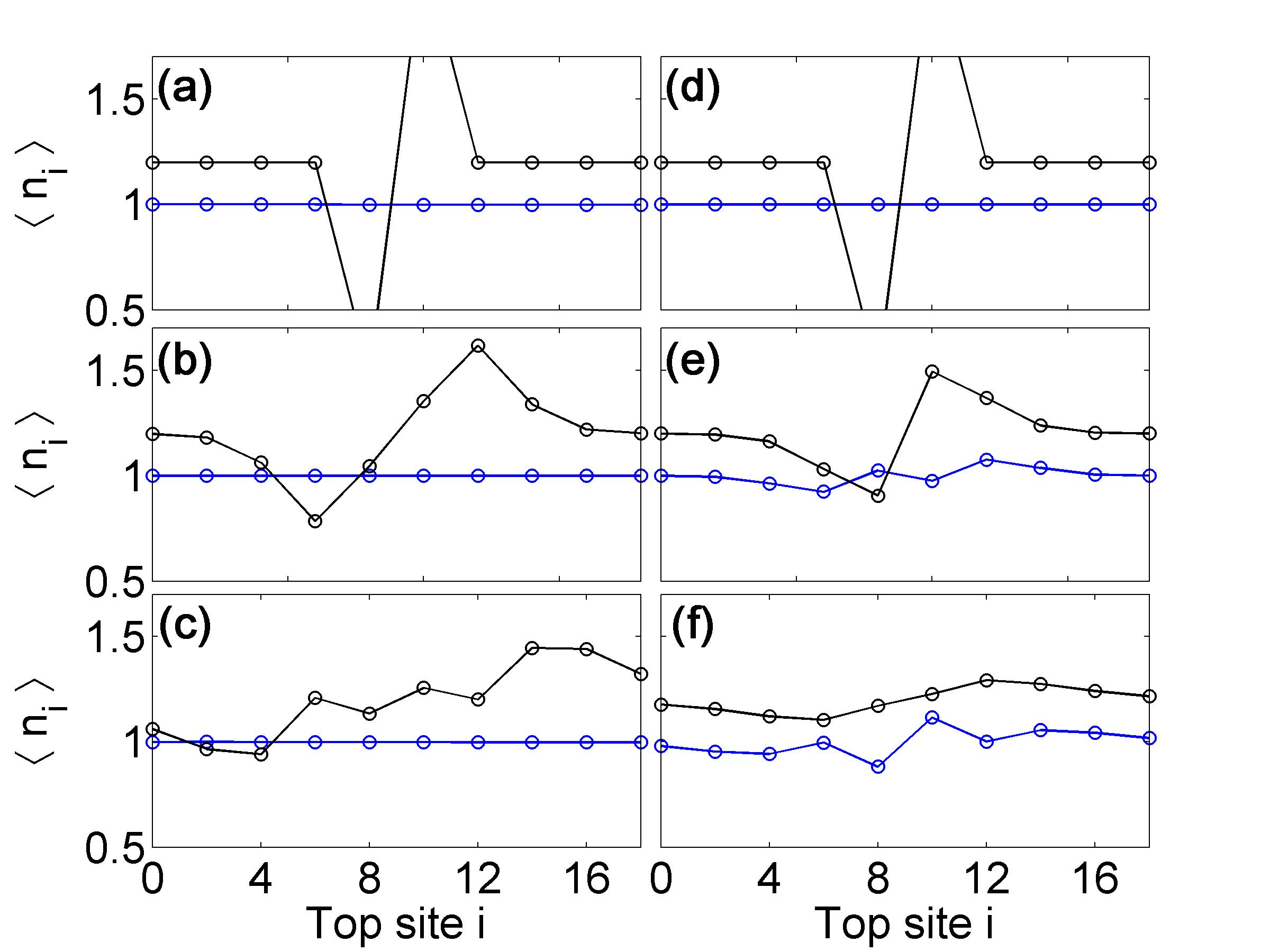}
\caption{\label{fig:ChargeAllSites} (Color online) Charge $\langle n_i \rangle$
at each site for different times for the cases $t_y=0$ (a,b,c) and $t_y=3$
(d,e,f). Charge values in the upper leg are shifted by $0.2$ for better
visualization. Times are $\tau=0$ (panels a,d), $\tau=1$ (b,e), and $\tau=2$
(c,f). }
\end{figure}

The typical behavior of the charge $\langle n_{i}\rangle(\tau)$ as a function
of time is shown in Fig.~\ref{fig:ChargeAllSites} for $U=10$. Each panel shows
the charge configuration of the ladder at a given time $\tau$ for $t_y=0$ (left
panels) and $t_y=3$ (right panels). The top panels show the creation of the
doublon $(\langle n \rangle =2)$ and holon $(\langle n \rangle =0)$ excitations
in the central sites of the upper leg at $\tau=0^+$. As time progresses, the
excitations propagate throughout the system in opposite directions,
\cite{Al-Hassanieh:166403:2008} eventually reaching the end of the ladder at
$\tau\sim3-4$.

\begin{figure}[t]
\includegraphics[width=1.0\columnwidth,clip]{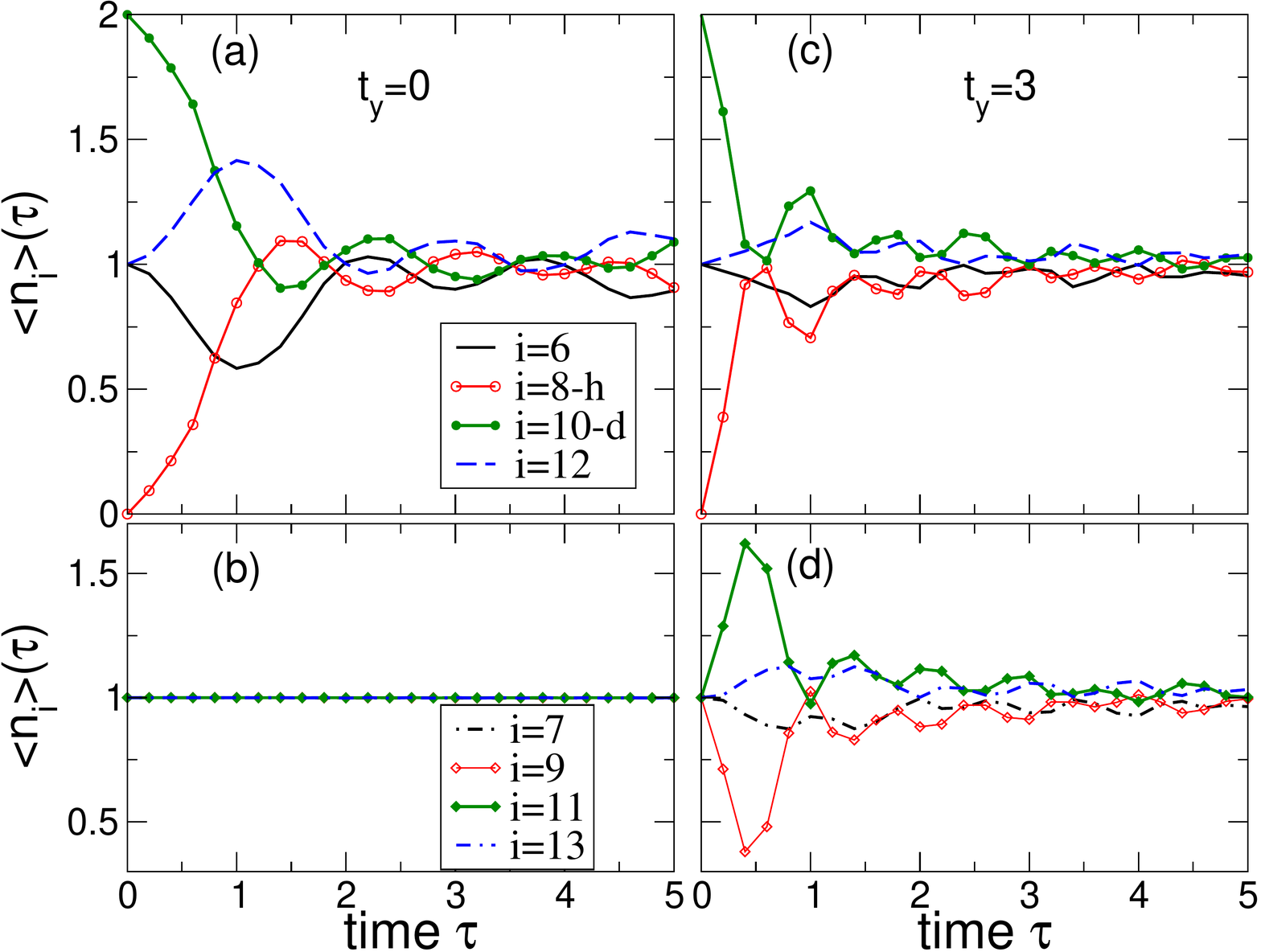}
\caption{\label{fig:ChargeSite} (Color online) Charge $\langle n_i \rangle$ on
sites close to the holon-doublon creation sites (8 and 10) for $t_y=0$ (a,b),
and $t_y=3$ (c,d).}
\end{figure}

More instructive is to show the average charge as a function of time $\tau$ on
a particular site. Fig.~\ref{fig:ChargeSite} shows  $\langle n_i \rangle(\tau)$
at the sites close to the holon-doublon creation point for $U=10$, $t_x=1$ and
$t_y=0$ (corresponding to the case of uncoupled ladder legs, with zero
spin-gap) and $t_y=3$ (spin gapped case).

For $t_y=0$ (left panels in Fig.~\ref{fig:ChargeSite}) and $\tau=0$, the charge
at holon and doublon sites is 0 and 2, respectively with $\langle n_i
\rangle=1$ in the remaining sites. As time evolves, the charge on the doublon
(holon) site decreases (increases) and oscillates  as time progresses. More
importantly, the excess (missing) charge is initially transferred to the
neighboring sites in the same leg (labeled ``6" and ``12") so that the charge
excitations ``move" throughout the upper leg and, eventually, the local charge
on all sites in the upper leg will fluctuate, while keeping the total charge
$N=\sum_i \langle n_i \rangle$ constant. As expected, for $t_y=0$ there is no
transfer to the lower leg (Fig~\ref{fig:ChargeSite}-b).

For $t_y \neq 0$, the charge excitations will also propagate into the lower leg
(Fig.~\ref{fig:ChargeSite}-c,d). In the case $t_y=3 t_x$, the dynamics of the
local charge on the sites neighboring, say, the doublon site (site ``10")
vertically or horizontally will be different, with a large portion of charge
oscillating vertically due to the dominance of the hoppings along the $y$
direction ($t_y>t_x$). This can be seen in Fig.~\ref{fig:ChargeSite}(c,d) by
comparing the charge on sites ``11" (first neighbor in the lower leg) and ``12"
(first neighbor in the upper leg).
%

We should also point out that, since the time-evolution conserves energy, there
is no mechanism for ``recombination" of the doublon-holon pair once it is
created.\cite{Al-Hassanieh:166403:2008} Thus, the electron- and hole-like
excitations do not ``cross" each other, creating a spatial particle-hole
asymmetry in the system. This asymmetry is kept throughout the time evolution
such that the charge of sites in the ``electron side" are related to their
``hole side" counterparts by a particle-hole transformation (e.g., $\langle
n_{i=12} \rangle(\tau)=2-\langle n_{i=6} \rangle(\tau)$).

\subsection{Double Occupancy}
\label{sec:doubleoccupancy}

\begin{figure}[t]
\includegraphics[width=1.0\columnwidth,clip]{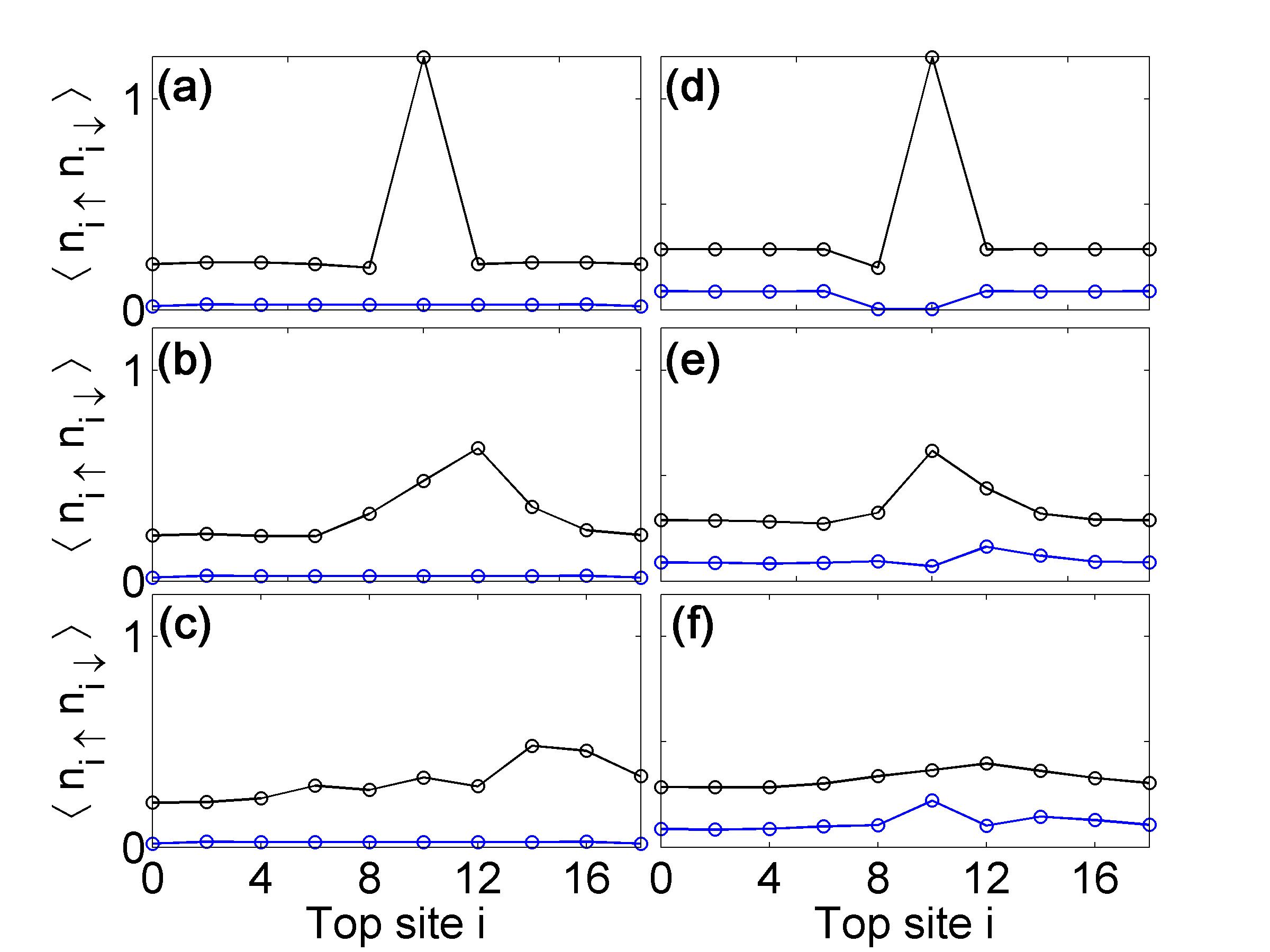}
\caption{\label{fig:DoccAllSites} (Color online) Double occupancy $\langle n_{i
\uparrow} n_{i \downarrow} \rangle$ at each site for different times for the
cases $t_y=0$ (a,b,c) and $t_y=3$ (d,e,f). Double occupancy values in the upper
leg are shifted by $0.2$ for better visualization. Times are $\tau=0$(a,d),
$1$(b,e), and $2$(c,f). }
\end{figure}

\begin{figure}[t]
\includegraphics[width=1.0\columnwidth,clip]{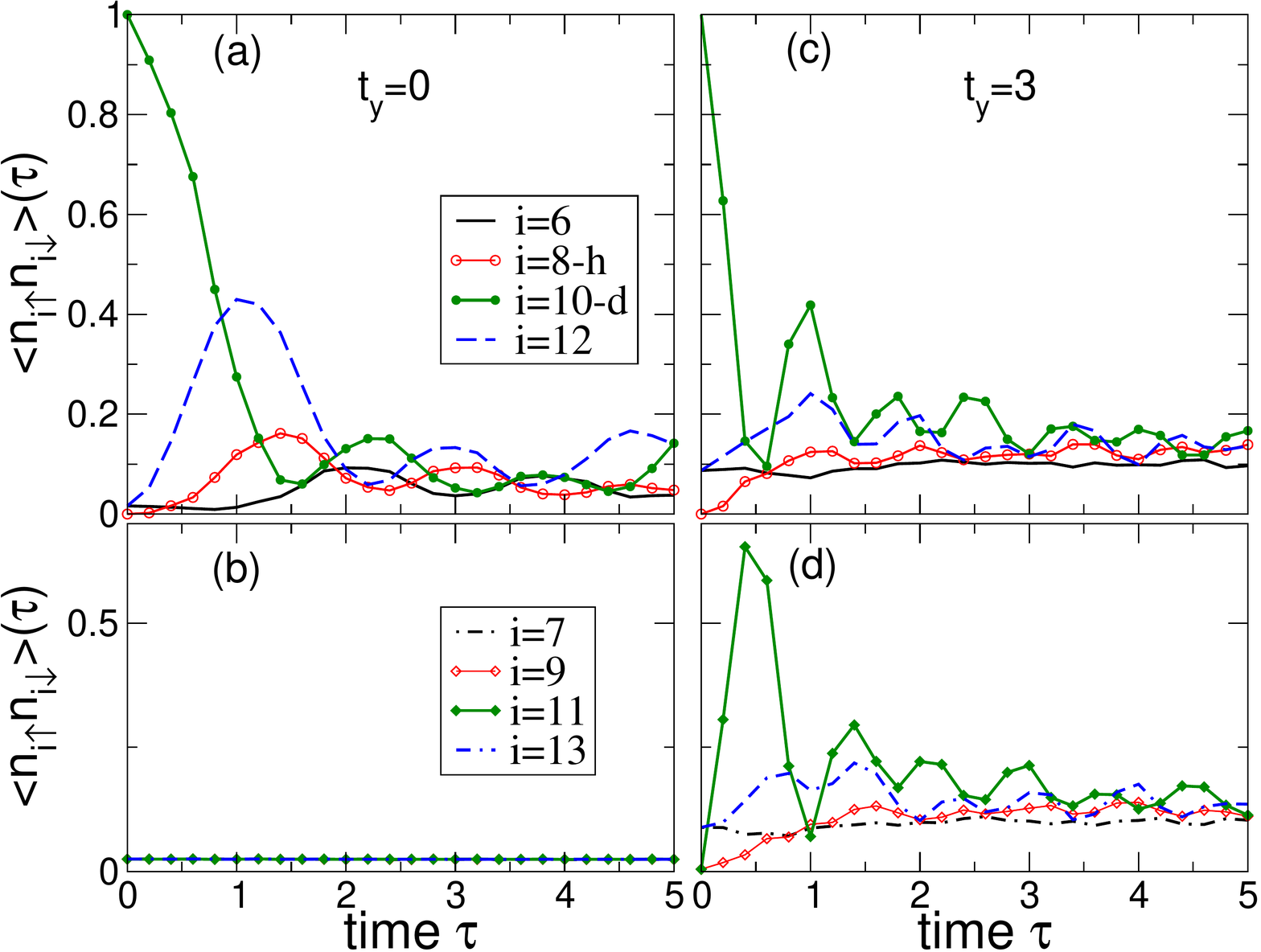}
\caption{\label{fig:DoccSite} (Color online) Double occupancy versus time at
sites close to the holon-doublon creation sites (8 and 10) for $t_y=0$ (a,b)
and $t_y=3$ (c,d).}
\end{figure}

The behavior of the double occupancy $\langle n^d_{i} \rangle \equiv \langle
\hat{n}_{i \uparrow}\hat{n}_{i \downarrow} \rangle$ is shown in
Figs.~\ref{fig:DoccAllSites} and \ref{fig:DoccSite}. At $\tau=0^+$, the double
occupancy is the maximum ($=1$) at the doublon site, zero at the holon site and
small in the rest of the system (the actual value will depend on $U$). For
$t_y=0$, the local double occupancy fluctuates as the doublon-holon pair
propagates in the upper leg but there is no change in the double occupancy in
the lower leg. This is in contrast with the behavior for $t_y > t_x$, for which
the propagation of the doublon in the upper leg changes the double occupancy of
the sites in the lower leg.

An interesting property seen for $t_y \neq 0$ is that the vertical coupling
correlates the double occupancy of sites within the same ``rung".  This is
clearly shown in Fig.~\ref{fig:DoccAllSites}-d, which shows the double
occupancy at $\tau=0^+ $ (compare it with the $t_y=0$ case
Fig.~\ref{fig:DoccAllSites}-a): the creation of a holon or a doublon on a given
site also changes the double occupancy in the site directly below it. This
happens even in the non-interacting case and it is a natural consequence of the
dominance of the rungs in the $t_y \gg t_x$ regime.

This effect can be better visualized by looking at the double occupancy for
individual sites as a function of time (Fig.~\ref{fig:DoccSite}). For $t_y=3$
(Fig.~\ref{fig:DoccSite}-c,d) and time $\tau=0^+$, when the holon-doublon pair
is created at sites 8 and 10, the double occupancy at sites in the same rungs
(9 and 11, respectively) is reduced to zero. This is in contrast with the other
sites in the lower leg (say, sites 7 and 13), which retain the ground-state
double occupancy value. This is quite different from the case of uncoupled legs
($t_y=0$, Fig.~\ref{fig:DoccSite}-a,b), for which no change in the double
occupancy happens in the lower leg, as expected.

This  ``non-local" change in the double occupancy in the lower leg when a
doublon is created  in the upper leg can be illustrated in a very simple model.
Let us consider a single, two-site rung (say, sites 0 and 1) with a hopping
term $H_{\rung}=t_y\sum_{\sigma} \hat{c}^{\dagger}_{0 \sigma}\hat{c}_{1
\sigma}+\mbox{h.c.}$ between them. The ground state $|\mbox{GS}\rangle$ of
$H_{\rung}$ lies in the two-electron sector (with energy $-2t_y$) and it is given by
\begin{equation}
\label{Eq:RungGS} |\mbox{GS}\rangle=\frac{1}{2}\left(-\sqrt{2}|S\rangle_{\rung}
+ |D_0\rangle + |D_1\rangle \right) \; ,
\end{equation}
where $|S\rangle_{\rung} = (1/\sqrt{2})
\left(|\uparrow\rangle_{0}|\downarrow\rangle_{1} -
|\downarrow\rangle_{0}|\uparrow\rangle_{1} \right)$ is a rung spin singlet
involving spins on sites 0 and 1 and
$|D_{0(1)}\rangle=|\uparrow\downarrow\rangle_{0(1)}|0\rangle_{1(0)}$ are doubly
occupied states. \cite{NoteUneq0}

Notice that the double occupancy is the \textit{same} in both sites:
$\langle\mbox{GS}| \hat{n}_{0(1)\uparrow}
\hat{n}_{0(1)\downarrow}|\mbox{GS}\rangle = 1/4$. Now, if a doublon is created
at site 0, then the (normalized) state becomes:
\begin{equation}
\label{Eq:RungState} |\psi_{d0}\rangle=d^{\dagger}_{0} |\mbox{GS}\rangle =
|\uparrow\downarrow\rangle_{0}\otimes\frac{\left(|\uparrow\rangle_{1} +
|\downarrow\rangle_{1}\right)}{\sqrt{2}}  \; ,
\end{equation}
and, obviously, the double occupation in site 0 is now equal to $1$. Notice,
however that the double occupancy in site 1 will be \textit{reduced}, since
$\langle\psi_{d0}|\hat{n}_{1\uparrow} \hat{n}_{1\downarrow}|\psi_{d0}\rangle =
0$, hence a ``non-local" change in double occupancy occur at site 1 as the
doublon is created in site 0.

Another interesting feature of the large $t_y$ regime is the ``beating" of both
charge and double occupancy within the rung where the doublon is created.
Figs.~\ref{fig:ChargeSite}-(c,d) and ~\ref{fig:DoccSite}-(c,d) show that, as
time evolves, the charge and double occupancy in site 11 oscillate with the
same period but out of phase with their counterparts in site 10. Interestingly,
for the ``holon rung" (sites 8 and 9) these beatings show up only in the charge
(Fig.~\ref{fig:ChargeSite}-c,d) but not in the double occupancy
(Fig.~\ref{fig:DoccSite}-c,d).

Next, we look at the total double occupancy $N_d(\tau)=\sum_i \langle n^d_i
\rangle(\tau)$. In contrast with the total charge (which is conserved and
therefore independent of $\tau$), $N_d(\tau)$ fluctuates with time $\tau$ for
$U\neq0$. As in Ref. \onlinecite{Al-Hassanieh:166403:2008}, we look at the
quantity $\Delta N_d (\tau)=N_d(\tau)-N^{(0)}_d$ where $N^{(0)}_d =\langle
\Psi_0 | n^d_i | \Psi_0 \rangle$ is the total ground-state double occupancy.
This ``excess double occupancy" is shown in Fig.~\ref{fig:DoubleOccDecay}.

We note that the value of $\Delta N_d (0)$ will depend on $t_y$. For $t_y=0$,
the creation of the holon-doublon pair only changes the double occupancy
locally and therefore $\Delta N_d (0)=1-\left(\langle \Psi_0 | n^d_{ih} |
\Psi_0 \rangle+\langle \Psi_0 | n^d_{id} | \Psi_0 \rangle\right)$ where $ih$
and $id$ label the sites where the holon and doublon are created, respectively.
By contrast, for $t_y \neq 0$, the creation of a doublon-holon pair will change
the double occupancy on the other sites of the rungs as well (as discussed
above) by an amount that increases with $t_y$. Thus, $\Delta N_d (0)$ decreases
with $t_y$, as shown in the inset in Fig.~\ref{fig:DoubleOccDecay}. The
decrease is much accentuated for $t_y\geq2$, establishing a clear qualitative
distinction from the results for the one-dimensional case $t_y=0$, where
$\Delta N_d (0)$ is ``local" in the sense that it depends only on the
ground-state double occupancy at the sites where the holon-doublon is
created.\cite{Al-Hassanieh:166403:2008}

For longer times, $\Delta N_d (\tau)$ reaches a plateau with a height that also
decreases with $t_y$. For $U=10$ and large $t_y$, $\Delta N_d (\tau)$
fluctuates around a value that is essentially independent of $\tau$, with no
sizeable decay in double occupancy with time. This is shown in the inset of
Fig.~\ref{fig:DoubleOccDecay} where $\Delta N_d (\tau=0^+)$ and $\Delta N_d
(\tau=6)$ are plotted as a function of $t_y$. For $t_y=0$ and different values
of $U$, the value of $\langle \Delta N_d \rangle_{\tau}$ increases with $U$,
consistently with the findings of Ref. ~\onlinecite{Al-Hassanieh:166403:2008}.

\subsection{Doublon speed}
\label{sec:doublonspeed}

\begin{figure}[t!]
\includegraphics[width=1.0\columnwidth,clip]{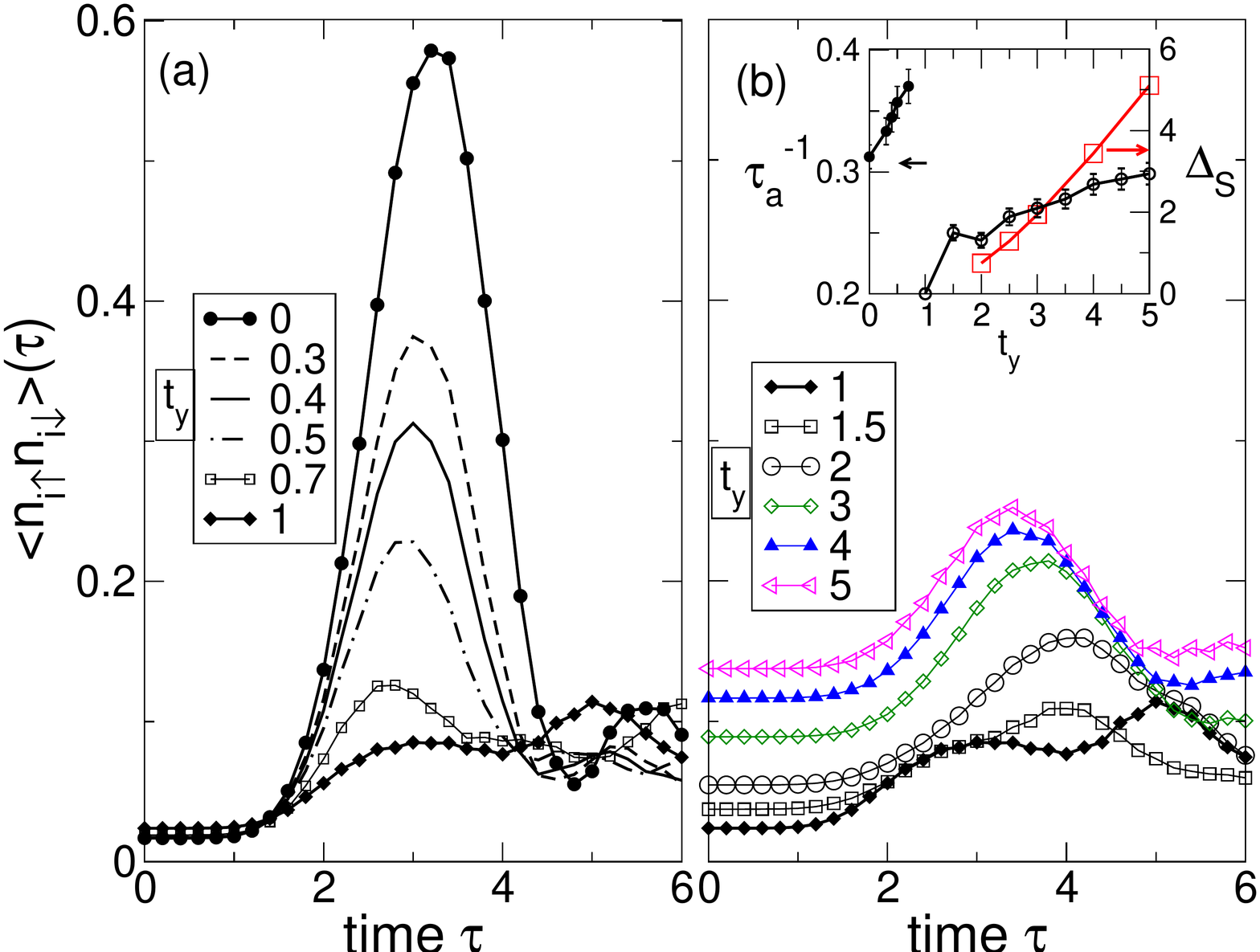}
\caption{\label{fig:DoccSite18} (Color online) Double occupancy at a site at
the end of the ladder for (a) $t_y\leq1$ and (b) $t_y\geq1$ (no shift added).
The position of the peak marks the ``arrival time" $\tau_a$ of the doublon.
Inset: inverse doublon arrival time $\tau^{-1}_a$ (circles) and spin-gap
calculated for a $2\times32$ ladder (squares) vs $t_y$. Error bars are given by
$\Delta \tau/\tau^2$ and indicate the uncertainty in the determination of the
arrival time given the time step $\Delta \tau$.}
\end{figure}

A clear difference in behavior between the two regimes (small and large $t_y$)
can be captured by the speed at which the charge excitations travel,
particularly the doublon. This is illustrated in Fig.~\ref{fig:DoccSite18}
which shows the double occupancy vs time at site ``18", located in the upper
leg and at the right end of the ladder.
\begin{figure}[t]
\includegraphics[width=1.0\columnwidth,clip]{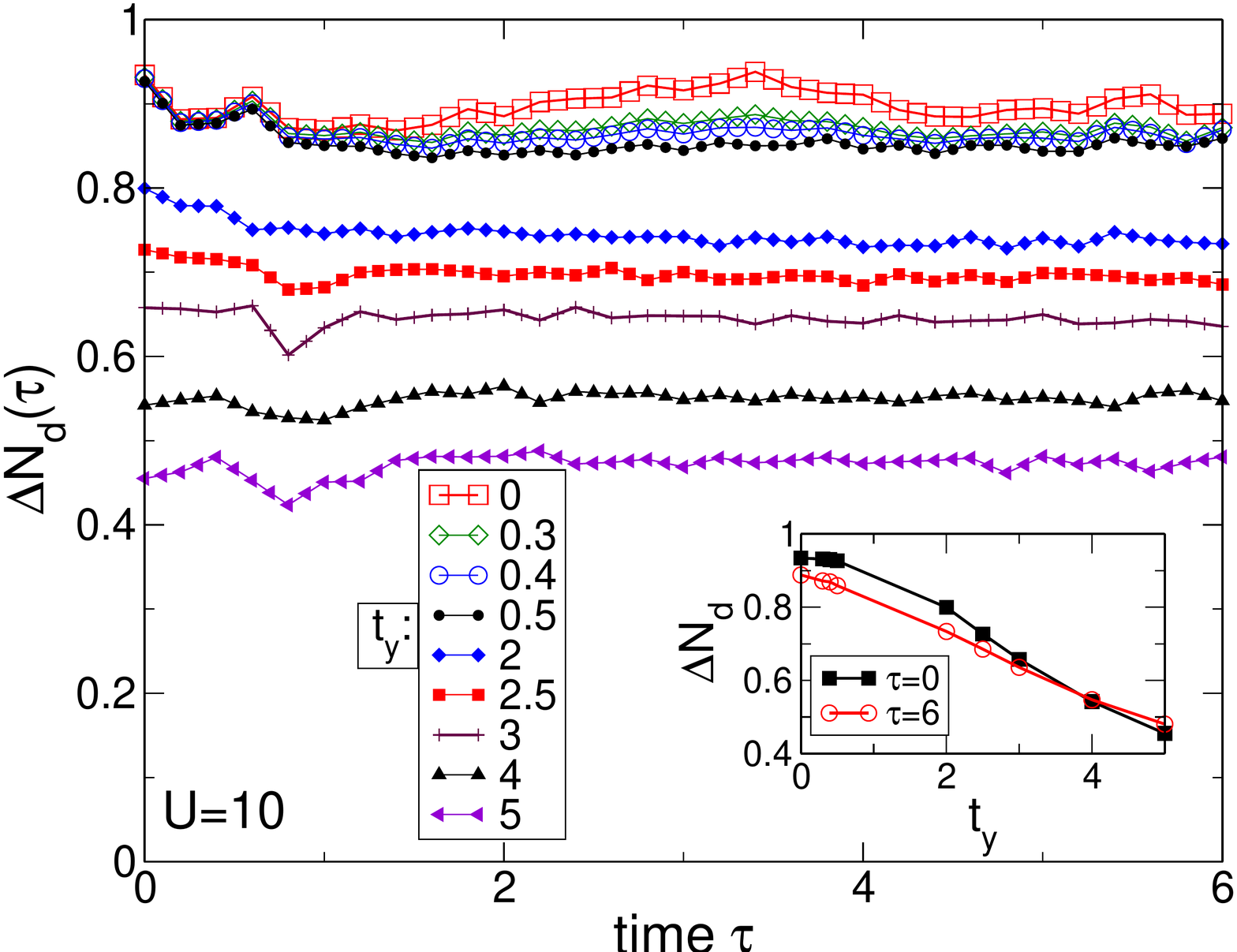}
\caption{\label{fig:DoubleOccDecay} (Color online) Excess double occupancy
$\Delta N_d (\tau)$ versus $\tau$ for $U=10$ and different values of $t_y$.
Inset: $\Delta N_d$ at fixed times ($\tau=0$ and $\tau=6$) versus $t_y$.}
\end{figure}

The position of the first peak in $\langle n_{d i} \rangle(\tau)$ marks the
``arrival time" $\tau_a$ of the doublon and can be used to infer its ``speed"
$v_d \propto \tau^{-1}_a$. In Fig.~\ref{fig:DoccSite18}, we can identify two
distinct regimes: for $t_y\leq1$ (Fig.~\ref{fig:DoccSite18}-a), such arrival
times are of order $\tau_a \sim 3$ while for $t_y>1$
(Fig.~\ref{fig:DoccSite18}-b), arrival times are substantially higher, in the
$3 < \tau_a \lesssim 4$ range. Thus, the doublon effectively moves ``faster"
(smaller arrival times) for small values of the inter-leg coupling $t_y$.

The clear difference in arrival times in the two regimes is striking. Some
hints as to why it comes about can be found in the curve for $t_y/t_x \sim 1$.
In the transition between the two regimes ($t_y/t_x \sim 1$), a second peak,
with a longer arrival time, becomes more prominent than the first peak. As
$t_y$ increases, this second peak dominates and effectively marks the arrival
of the doublon. In fact, this second peak arises as a result of the part of the
doublon that travels through the \textit{lower leg} instead of the upper one
for large $t_y$, as shown in Figs.~\ref{fig:DoccAllSites} and
\ref{fig:DoccSite}.

A curious behavior is that, within both regimes ($t_y \leq 0.5$ and $t_y \geq
2$), the horizontal doublon speed increases with the coupling between the
ladder legs. This is shown in the inset, where $\tau^{-1}_a$  is plotted as a
function of $t_y$ (circles). We notice that, within each regime, $\tau^{-1}_a$
(taken from the more prominent peak at each regime, except for $t_y=1$ where
two values are used) does not depend linearly with $t_y$, indicating that such
increase does not arise from a simple rescaling of the ``time unit" in terms of
$\hbar/t_y$.

In addition, the scaling of $\tau^{-1}_a$ with $t_y$ is different in each
regime and therefore one might speculate if there is an extra energy scale
dominating the behavior of $\tau^{-1}_a$ at large $t_y$. As it is known from
previous studies,\cite{Noack:PhysicaC:281:1996} Hubbard two-leg ladders display
a sizeable spin gap in the large $t_y$ regime. One possibility would be that
the propagation of the doublon would be coupled to $S=1$ spin excitations
(where a singlet in a rung becomes a triplet), thus resulting in a scaling of
the doublon speed with the spin gap.

In order to check for such a connection between $\tau^{-1}_a$ and the spin gap
$\Delta_s$, we have calculated $\Delta_s$ for $t_y/t_x \geq2$. The spin gap can
be calculated with DMRG as
\begin{equation}
\Delta_s=E(N/2+1,N/2-1)-E(N/2,N/2)
\end{equation}
where $E(N_\uparrow,N_\downarrow)$ is the energy of a state with $N_\uparrow$
electrons with spin up and $N_\downarrow$ electrons with spin down (total
$N=N_\uparrow+N_\downarrow$ electrons). We have performed these (static)
calculations for a $2\times32$ ($N=64$) ladder with $U=10$ at half-filling.

The results are shown in the inset of Fig.~\ref{fig:DoccSite18}. Although the
results do not show a clear scaling of the type ``$\tau^{-1}_a \propto
\Delta_s$", the dependence of $\tau^{-1}_a$ with $t_y$ is markedly different in
the strong spin-gap regime as compared to ``small $t_y$" one: the rate of
increase of the doublon speed is lower for $t_y >2t_x$, where a clear spin gap
is present. This indicates that the propagation of the doublon might be coupled
to magnon-like spin excitations, although this connection is weak. This is also
consistent with the rather weak decay shown in Fig.~\ref{fig:DoubleOccDecay}
even in the large $t_y$ regime.

\section{Conclusions}
\label{sec:conclusion}

In this work, we have investigated the dynamics of holon-doublon pairs in
Hubbard two-leg ladders using the time-dependent DMRG method. The geometry of
the ladder brings some interesting qualitative changes as compared to the case
of the chains. A telling example is a ``non-local" character of the double
occupancy: the creation of a doublon in a site at the upper leg
\textit{reduces} the double occupancy in the bottom site coupled to it. This,
in fact, reduces the total double occupancy as the inter-leg coupling
increases.

In addition, we have found important qualitative differences in the dynamics of
the excitations depending on the relative coupling between the ladder legs. For
weak inter-leg coupling ($t_y/t_x \ll 1$), results are qualitatively similar to
the strictly one-dimensional
case.\cite{Al-Hassanieh:166403:2008,DiasdaSilva:125113:2010}. However, the
results show a strong downward shift in the doublon horizontal speed as one
crosses over from weak ($t_y/t_x \ll 1$) to strong ($t_y/t_x \gg 1$) inter-leg
coupling. We attribute this shift to different propagation ``paths" for the
doublon: for small $t_y$, the doublon propagates mainly through the sites of a
single leg. For large $t_y$, the propagation involves also the vertical
direction (through the ladder rungs rather than single sites), effectively
``slowing down" the doublon.

An intuitive physical picture of the doublon propagation in the $t_y\gg t_x$
regime emerges by noticing that, in this regime, it is often  a good
approximation
\cite{Martins:Phys.Rev.Lett.:3563:1997,Laukamp:Phys.Rev.B:10755:1998}  to
assume that the ground state of the ladder system is made out of individual
rung states, such as the one given by Eq.~\ref{Eq:RungGS}. For strongly
interacting ladders, the rung ground state has a strong contribution from the
spin singlet $|S\rangle_{\rung}$.\cite{NoteUneq0} Thus, the excitations
propagate in a ``background" of spins tightly packed into singlets. The
propagation then ``damages the background", causing a reduction of the
excitation's speed. We note that this is in sharp contrast with the
one-dimensional case, where no ``spin-singlet background" is present.
%

Moreover, although in both regimes ($t_y/t_x \ll 1$ and $t_y/t_x \gg 1$) the
``horizontal" speed of the doublon increases as the vertical coupling
increases, the scaling of the doublon speed with $t_y$ is different, being less
pronounced in the strong inter-leg coupling regime. Our results indicate that
the presence of a spin gap qualitatively changes such scaling, indicating a
possible (although weak) coupling between the doublon and spin excitations.

\begin{acknowledgments}
We thank Khaled Al-Hassanieh and Adrian Feiguin for fruitful discussions. This
work was supported by the Center for Nanophase Materials Sciences, sponsored by
the Scientific User Facilities Division, Basic Energy Sciences, U.S. Department
of Energy, under contract with UT-Battelle. This research used resources of the
National Center for Computational Sciences, as well as the OIC at Oak Ridge
National Laboratory. L.D.S acknowledges support from Brazilian agencies CNPq
and FAPESP (Grant No. 2010/20804-9). G.A. acknowledges support from the DOE
early career research program. E.D. is supported in part by the U.S. Department
of Energy, Office of Basic Energy Sciences, Materials Science and Engineering
Division.
\end{acknowledgments}

\appendix

\section{Accuracy during time evolution}
\label{sec:TechnicalComments}

In this section, we discuss some technical aspects regarding the accuracy of
the results.
%
%
In addition to the usual criteria for detecting loss of accuracy during time
evolution (namely, an exponential increase of the truncation error for a given
number $m$ of kept states, leading to a ``runaway time"
\cite{Gobert:Phys.Rev.E:036102:2005}), we have also monitored the energy
expectation value $E(\tau) \equiv \langle\Psi (\tau)|\hat{H}|\Psi
(\tau)\rangle/\langle\Psi (\tau)|\Psi (\tau)\rangle$ with increasing $\tau$ as
a way to probe the accuracy of the results for longer time scales. Since the
Hamiltonian in Eq.~\ref{Eq:Hamiltonian} is time-independent, the expectation is
that this quantity should be constant (i.e., $E(\tau)=E(0)$) and thus
deviations from $E(\tau)=E(0)$ can be used as a gauge of the accumulation of
numerical errors during the time evolution of the system.
\begin{figure}[t]
\includegraphics[width=1.0\columnwidth,clip]{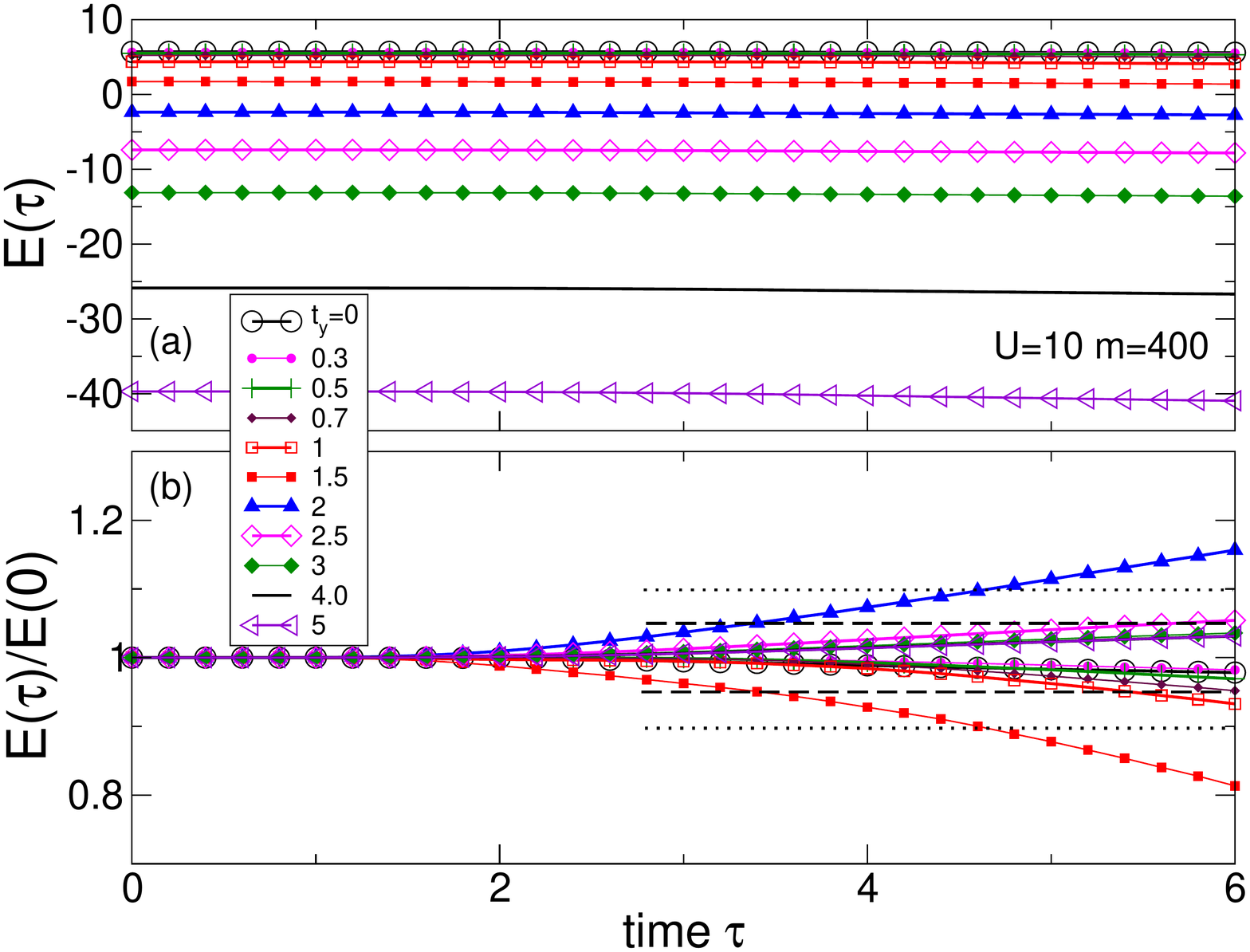}
\caption{\label{fig:EnTime} (Color online) (a) Average energy $E(\tau)$ as a
function of $\tau$ for $U=10$ with $m=400$ kept states and different values of
$t_y$. In (b), $E(\tau)/E(0)$ is plotted, with dashed and
dotted horizontal lines marking the 5\% and 10\% ``confidence ranges"
respectively.
}
\end{figure}

The results for $E(\tau)$ are shown in Fig.~\ref{fig:EnTime}. As time
progresses, the accumulated errors during the time evolution cause $E(\tau)$ to
deviate from the value at $\tau=0$.
Since we have opted to keep $U$ and $t_x$ fixed (no reescaling),
the numerical value of  $E(0)$ (the average energy of the system after the
creation of the holon-doublon pair) will change significantly as $t_y$ is
increased (Fig.~\ref{fig:EnTime}-a). For better visualization, we have also
plotted $E(\tau)/E(0)$ (Fig.~\ref{fig:EnTime}-b), which gives a measure of the
relative ``degradation" (i.e., the deviation from $E(\tau)/E(0)=1$). As seen in
Fig.~\ref{fig:EnTime}-b,
this ``degradation" in energy will increase with time but it is within 5\%
accuracy up $\tau \sim 6$, using $m=400$ states during the time evolution and
time step $\Delta \tau=0.2$, which we consider acceptable.

We should note that, as can be inferred from Fig.~\ref{fig:EnTime}-a, $E(0)$
passes through zero and becomes negative for $t_y\sim 1.75$, causing
$E(\tau)/E(0)$ to be larger than 1 for $t_y\gtrsim 1.75$ although
$E(\tau)<E(0)$ in general (i.e., the average energy always \textit{decreases}).
Additionally, the fact that $E(0)$ is relatively small ($|E(0)|<5$) for $t_y$
in the range $1.5-2$ makes the deviation from $E(\tau)/E(0)=1$ to be larger
(see Fig.~\ref{fig:EnTime}-b), although this does not imply a larger
accumulated error (in fact, in this $t_y$ range, $|E(\tau)-E(0)|<0.4$ up to
$\tau \sim 6$).

%

\begin{figure}[t]
\includegraphics[width=1.0\columnwidth,clip]{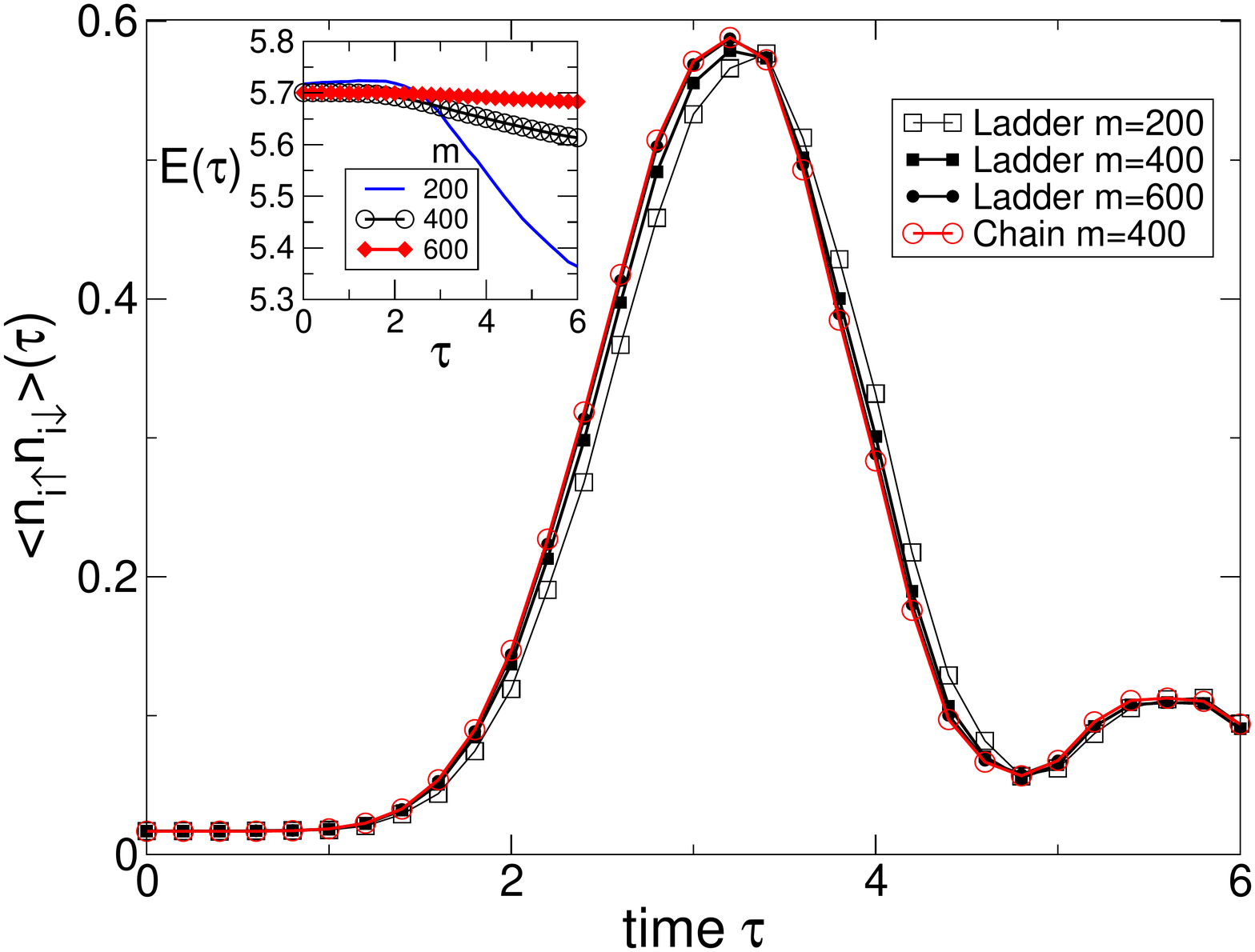}
\caption{\label{fig:CompLadderChain} (Color online) Comparison between
ladder and chain calculations: double occupancy at a site located near the end
of the system vs time  for a chain with $m=400$
kept states and a ladder with $t_y=0$ and $m=200,400$ (same data as in Fig.~\ref{fig:DoccSite18}-a) and $600$ kept states.
Inset: $E(\tau)$ in the ladder for $t_y=0$ and different number of kept states
during the time evolution.}
\end{figure}

Another important accuracy test is a comparison between the propagation of the
holon-doublon in an uncoupled ladder ($t_y=0$) and in a one-dimensional chain.
These should yield similar results if the number of kept states $m$ is
sufficiently large. As shown in Fig.~\ref{fig:CompLadderChain}, we find that
using $m=400$ kept states in the ladder reproduces the chain results for the
double occupancy quite well at a quantitative level up to $\tau \sim 6$. In
fact, even a much smaller number of kept states $m=200$ already give the
correct qualitative behavior. Moreover, we find that increasing $m$ above
$m=400$ does not modify the trends and/or numerical values appreciably: the
results for the time-dependence of observables such as double occupancy are
essentially unchanged up to $\tau \sim 6$.

Thus, the results presented in the paper (e.g., the doublon speed in
Fig.~\ref{fig:DoccSite18}) calculated with $m=400$ will not be significantly
modified by increasing $m$. Keeping more states (say, $m=600$) during time
evolution brings only a slight improvement in both the observables and also in
$E(\tau)$. This latter point is illustrated in the inset of
Fig.~\ref{fig:CompLadderChain}, where $E(\tau)$ is shown for $t_y=0$ and
different values of $m$. For this reason, the calculations presented in this
paper were performed keeping up to $m=400$ states during the time evolution, as
increasing the number of retained states to $m=600$ does not justify the
considerable extra computational cost.


\end{document}